%% file: graphene.tex
\documentclass[twocolumn,letterpaper,superscriptaddress,showpacs,floatfix,prl]{revtex4}
\usepackage{natbib,hyperref}
\usepackage{epsfig}
\usepackage[utf8]{inputenc}
\usepackage{color}
\usepackage{amsmath}
\usepackage{hyperref}   

\begin{document}
\pagestyle{plain}
\newcount\eLiNe\eLiNe=\inputlineno\advance\eLiNe by -1
\title {Van der Waals interlayer potential of graphitic structures: from Lennard-Jones to Kolmogorov-Crespy and Lebedeva models.}
\author{Zbigniew Kozioł\footnote{e-mail: zbigniew.koziol@ncbj.gov.pl}}
\affiliation{National Center for Nuclear Research, Materials Research Laboratory, ul. Andrzeja Sołtana 7, 05-400 Otwock-Świerk, Poland}
\author{Grzegorz Gawlik}
\affiliation{Institute of Electronic Materials Technology, 133 Wólczyńska Str.,01-919 Warszawa, Poland}
\author{Jacek Jagielski}
\affiliation{National Center for Nuclear Research, Materials Research Laboratory, ul. Andrzeja Sołtana 7, 05-400 Otwock-Świerk, Poland}
\affiliation{Institute of Electronic Materials Technology, 133 Wólczyńska Str.,01-919 Warszawa, Poland}

\begin{abstract}
The experimental knowledge on interlayer potential of graphenites is summarized and compared with computational results based on phenomenological models. Besides Lennard-Jones approximation, the Mie potential is discussed, Kolmogorov-Crespy model and equation of Lebedeva et al. An agreement is found between a set of reported physical properties of graphite (compressibility along c-axis under broad pressure range, Raman frequencies for bulk shear and breathing modes under pressure, layer binding energies), when a proper choice of model parameters is made. It is argued that the Kolmogorov-Crespy potential is the preferable one for modelling. A simple method of fast numerical modelling, convenient for accurate estimation of all these discussed physical properties is proposed. It is useful in studies of other van der Waals homo/heterostructures.
\end{abstract}


\maketitle 

\input{introduction.tex}


\input{analytical.tex}



\section{Summary and Conclusions.}

Interlayer interaction modeling in graphenites based 
on phenomenological van der Waals potentials (Lennard-Jones, Kolmogorov-Crespi and Lebedeva’s was performed. 
Results have been compared self-consistently with ab-initio calculations and experimental data on 
compressibility and Raman shifts in LBM and shear modes under pressure, favoring strongly anisotropic
Kolmogorov-Crespi model. Computation was done by using molecular dynamics package LAMMPS and a proposed
convenient, extendable scheme of computation suitable for fast numerical modeling of several 
physical quantities. The method is useful for studying other 2-dimensional 
homo- and heterostructures with van der Waals type interaction between layers.
It is argued that the value of the known Raman shift in shear mode is consistent with the difference in energy 
between AA and AB stacking of around 5 meV per atom. The models do not provide explanation for the reported
low content of ABC stacking in natural graphite.

\bibliographystyle{unsrtnat}
\bibliography{graphene}

\end{document}

%% file: introduction.tex

\section{Introduction.}


There is recently a growing interest in artificial heterostructures formed by 2-dimensional layers of graphene and other newly discovered materials, like these of  transition metal dichalcogenides (TMDCs), and named 'van der Waals heterostructures' \cite{Geim}. The reason of that is reachness of physical phenomena observed and high potential of their possible applications \cite{Nam}. Moiré patterns result when two layers are rotated \cite{Wijk}, \cite{Wijk2} and are related to van Hove singularities. Recently, in bilayer graphene twisted at 1.1 degree superconductivity was found with critical temperature of 1.7K \cite{YuanCao}. By manipulating doping levels these singularities are observed at angles up to 31 degrees \cite{HanPeng}. It was shown that free-standing vdW heterostructure 
of graphene on hexagonal boron nitride (hBN), where a small lattice mismatch exists shows a buckled atomic structure formed as Moiré
pattern \cite{Argentero}. There is already a number of laboratory prototypes of optoelectronic devices \cite{Lin}, like bolometers \cite{Skoblin}, photodetectos \cite{Wei}, \cite{muxiang}. Their characteristics can be tuned by elemental doping, surface chemical doping, intercalation and electrostatic gating. 

It is believed that artificial van der Waals heterostructures offer a broad field for research on novel materials for graphene/silicene, graphene/$MoS_2$, and silicene/$MoS_2$ systems \cite{Nam}, as well silicene \cite{Houssa} or $MoS_2$ alone \cite{Kvashnin}. Ab-initio calculations suggest the existence of magnetism in $ReS_2$ doped with Co, Fe, and Ni \cite{MLuo}. Formation of quantum-well Type-I heterostructure in van der Waals - interacting monolayers of $MoS_2$ and $ReS_2$ has been demonstrated \cite{Bellus} and Type-II band alignment was found in vdW heterojunction diodes based on $InSe$ and $GaS$ and graphene \cite{FYan}. $WSe_2$ and $MoS_2$ monolayers have been shown useful to create charge density modulation in electronic band valleys, resulting in valley polarization, and $Na$-doped $WS_2$ under strain is considered a candidate for spintronics \cite{LuoYin}. After spin injection from a ferromagnetic electrode transport of spin-polarized holes within the $WSe_2$ layer was observed \cite{OLS}. hBN and graphene vdW heterojunctions indicate on large potential of their use in spintronics \cite{Gurram}.  Light-induced negative differential transconductance phenomenon was realized on graphene/$WSe_2$ heterojunction transistor \cite{Shim}. DFT calculations show that vdW interactions dominate between antimonene and graphene layers and by applying electric field between them one can tune the height and type of Schottky contacts \cite{WeiLi}. Lattice dynamical theory and ab initio calculations indicate on the existence of piezoelectricity in 2D lattices comprising h-BN, $2H-MoS_2$, and other transition-metal dichalcogenides \cite{Michel}.

It is generally thought that van der Waals interaction, ubiquitous in nature, is caused by temporal fluctuations of electronic charge that induces dipoles of random amplitude \cite{Kawai}.  It plays a role in friction and adhesion between materials, absorption of molecules on surfaces. These relatively weak forces are responsive for binding between separate sheets of graphene.

While many electronic properties of graphene-based systems are known \cite{Rozhkov} still there are open questions concerning
the detailed form and physical mechanisms leading to inter-plane potential in these systems. There are two types of approaches used to describe the van der Waals potentials of graphitic systems: an ab-initio one, based on density functional theory, and the other uses empirical potentials \cite{Girifalco2}. 

The DFT calculations often give an underestimated values of  binding energy energy in case of van der Waals interactions
\cite{Charlier2}, \cite{Trickey}, \cite{Rydberg}, \cite{DiVincenzo}, \cite{Schabel} and 
energy difference between bindings of (preferred) AB and AA type of stacking, $E_{AB}$ and $E_{AA}$, which seems too large, as we will discuss. Recent diffusion quantum Monte Carlo calculations predict for instance $E_{AA}/E_{AB}$ of about 0.65 \cite{Mostaani}. Results however depend strongly on the functionals used to model vdW forces \cite{Birowska}, \cite{Chakarova}.

In this work we compare how well several properties of graphene/graphite can be described by using a few empirical potentials. Some of them are well based on DFT calculations. An advantage of using empirical potentials relies on ease of their implementation in any numerical methods. Moreover, we describe a scheme of computing several basic properties based on a simple analytical approach. The method may be easily extended to investigation of vdW homo- and heterostructures other than graphene/graphite. We demonstrate that by careful choosing of parameters of phenomenological potentials we may reproduce self-consistently experimental values of compressibility of graphite as well Raman shifts observed in shear mode under pressure and predict change of Raman shift under pressure for layer breathing mode (LBM).

\subsection{Lennard-Jones and Mie Potentials.}

An often used approximation of vdW potential is the 12-6 Lennard-Jones one, 
$U(r) = 4 \epsilon \left((\sigma/r)^{12} - (\sigma/r)^6\right)$, with two only adjustable parameters, where $r$ is distance between atoms.

It reproduces well interlayer distance and elastic constant of graphite. It was proved useful in describing several basic properties of $C_{60}$ molecules \cite{Ping}, with $\epsilon=2.964 meV$, $\sigma=3.407 \AA$.  It was used for modeling carbon nanotubes \cite{Jiang}, with $\epsilon=4.656 meV$ and $\sigma=3.825 {\AA}$. He et al. \cite{XQHe} obtain an analytical approximation for vdW forces acting between nanotubes. Authors extend their continuum model \cite{Kitipornchai} to study vibrations of multilayered graphene sheets. It is a convenient approximation in modeling reinforcement of composite materials with carbon nanotubes \cite{HTan}.

A more general form of vdW potential is one introduced by Mie  \cite{Mie}, \cite{Avendano} and it often reproduces well properties of carbon compounds, $U(r) = \alpha \epsilon \left(\left(\sigma/r\right)^m - \left(\sigma/r\right)^n\right)$, where $\alpha= \left( m/(m-n) \right) \cdot \left(m/n \right) ^ {n/(m-n)}$. When $m=12$ and $n=6$, the form of Lennard-Jones potential, with $\alpha=4$ is assumed. 


\subsection{Kolmogorov-Crespi and Lebedeva Potentials.}

\begin{figure}[ht]
\includegraphics[scale=0.9]{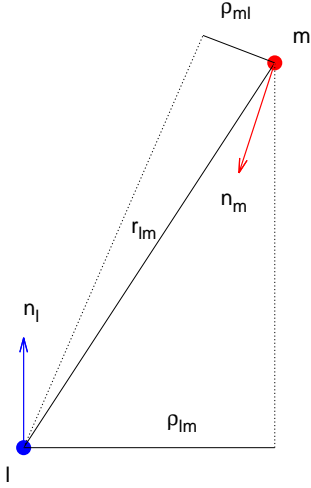}
      \caption{\label{KC00}Schematics of Kolmogorov-Crespi interaction between atoms {\bf l} and {\bf m} located on different flakes of graphene. The vector ${\bf n}_k$ (k=l,m) is normal to the $sp^2$ plane in the vicinity of atom k.
}
\end{figure}

There are at least two deficiencies of LJ or Mie type of potentials. First is that they are isotropic, i.e. the potential depends on distance between atoms, only. However, in graphenites, the binding between atoms on different planes is due to overlap between $\pi$ electrons
from adjacent layers and as such it ought to depend on the angle between orbitals. The second problem, as it was noticed first by Kolmogorov and Crespi \cite{Kolmogorov}, \cite{Kolmogorov2} and will also be shown in this work, is that these potentials produce too small energy difference between bindings of (preferred) AB and AA type of stacking, $E_{AB}$ and $E_{AA}$. Albeit there is some controversy about how large that energy difference is. The KC potential has an $r^{-6}$ two-body van der Waals-like attraction and an exponentially decaying repulsion terms, very short ranged, falling essentially to zero at two transverse interatomic distances . The directionality of the overlap is reflected by a function which rapidly decays with the transverse distance $\rho$ (Fig. \ref{KC00}). Most often it is used in the following form, for interaction between atoms $m$ and $l$:

\begin{widetext}
\begin{align}
U_{lm} & = -A\left(\frac{z_0}{r_{lm}}\right)^6 + exp \left(-\lambda(r_{lm}-z_0) \right) \cdot \left[C + f(\rho _{lm})+f(\rho _{ml}) \right] \\
\rho _{lm}^2 & = r_{lm}^2 - \left({\bf n}_l {\bf r} _{lm} \right)^2 \\
\rho _{ml}^2 & = r_{lm}^2 - \left({\bf n}_m {\bf r} _{lm} \right)^2 \\
f(\rho _{lm}) & = exp\left( -( \rho _{lm}/\delta )^2\right) \cdot     \sum _{n=0} ^{2}   C_{2n}     \left( \rho _{lm}/\delta \right) ^{2n}
\label{eq_guerra}
\end{align}
\end{widetext}

where ${\bf n}_k$ is the vector normal to the $sp^2$ plane in the vicinity of the atom $k$, and $z_0$ is close to  the
interlayer distance at equilibrium. The summation over $n$ in Eq. \ref{eq_guerra} is usually limited from $n=0$ to $n=2$ \cite{LAMMPS}.

KC potential has been used broadly in numerical modeling (molecular dynamics) \cite{Wijk}, \cite{Wijk2}, \cite{Wijk3},\cite{Annelot} as well in description of  ballistic nanofriction  \cite{Ng}, \cite{Guerra}.

Lebedeva et al. \cite{Lebedeva}, \cite{Lebedeva2} use another form of anisotropic potential which was obtained 
by fitting DFT calculations and was found to describe well the experimental graphite compressibility and 
the corrugation against sliding, while Jiang \cite{JinWuJiang} uses \ref{eq_lebedeva} for modeling thermal conductivity of FLG:

\begin{align}
U &= -A\left(\frac{z_0}{r}\right)^6 + B exp \left(-\alpha(r-z_0) \right) +\nonumber\\
  &+ C \left(1+D_1 \rho^2 + D_2 \rho^4  \right) exp \left( - \lambda _1 \rho ^2 \right) exp \left( - \lambda _2 \left( z^2 - z_0^2\right) \right)
\label{eq_lebedeva}
\end{align}

where $r$ is the interatomic distance and $\rho = \sqrt{r^{2}-z^{2}}$ is the projection of the distance within the graphene plane.

%% file: analytical.tex
\section{Analytical approximation.} 

Lattice spacing in graphene in hexagonal c-direction, is known well from X-ray diffraction, it is 3.3538 {\AA} at room temperature, and it does not change strongly with temperature \cite{Baskin}. The equilibrium interlayers spacing is 3.34 {\AA} in AB stacked \cite{Baskin} and 3.44 {\AA} in turbostratic graphene \cite{Kiang}. Neutron diffraction studies show that all in-plane C-C lengths are $1.422{\AA} \pm 0.001{\AA}$ (\cite{Trucano}).

\subsection{Distances between atoms in AB and AA stacking.}

There are two types of ordering of atoms on the nearest plane with respect to atoms on another plane. Let us call them type I and type II ordering. The type I results when atom is placed over another atom of the hexagon and type II  when atom is placed over the center of hexagon. Type I ordering is the only one in case of AA stacking, and in case of AB stacking equal number of atoms is in ordering of type I and II.

\begin{figure}[ht]
\centering
\includegraphics[scale=0.35]{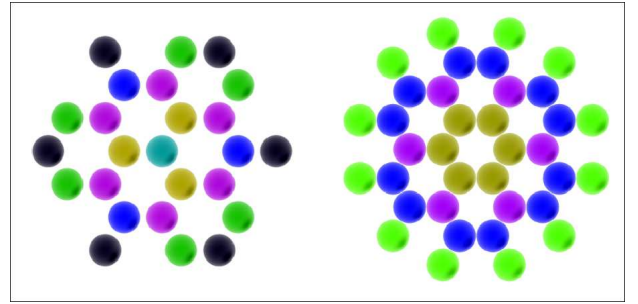}
      \caption{\label{ringsAAAB}
Rings of equi-distant atoms on neighbouring planes of graphene. On the left is type I, present for AA and AB stacking and type II (right) present in AB stacking, only.
}
\end{figure}

We can create a convenient scheme of calculating numerically potential energy for different stackings by computing 
it for each of these types of orderings. For that, we need distances projection (in plane) between an atom position over the plane together with number of atoms at these distances (rings of equi-distant atoms, as these in Fig. \ref{ringsAAAB}). These numbers can be found quickly with custom written scripts that compute distance between atoms in a graphene structure which was build also by scripts. The results are provided in Tables \ref{table_AA} and \ref{table_AB}, for both types of ordering. Analytical expressions on these distances are available however we do not know a method to express by formula the number of atoms in equi-distant rings.

In numerical simulations, e.g. performed in LAMMPS \cite{LAMMPS}, it is a standard procedure to introduce a cut-off distance 
in Equations \ref{MN_00} -- \ref{eq_lebedeva}. Potential energy between particles exceeding that distance is assumed to be equal zero, while the function given by these equations is appropriately smoothed out at that distance in order to avoid possible discontinuities in computational results. That cut-off is taken usually at around 12 {\AA} in case of graphene modelling. 

Limiting any approximation to atoms listed in Tables \ref{table_AA} and \ref{table_AB} is equivalent of using
cut-off distance in Eq. \ref{MN_00} of about 15.4 {\AA} and then we expect to obtain the same accuracy of results as by using LAMMPS.

\begin{widetext}

\begin{table}[h]
\caption{Distance between an atom on one graphene plane and nearest neighbors on next plane, for type I of neighbors, as in Fig. \ref{ringsAAAB} (left), found in AA and AB stacking of layers. \textbf{N} is number of neighbors and \textbf{d} is distance (in units of graphene unit cell value of 1.42  {\AA}).}
\label{table_AA}
\centering
\begin{tabular}{|c|cccccccccccccc|}
\hline
\textbf{d}	&0      &1      &1.7321 &2      &2.6458 &3      &3.4641 &3.6056 &4      &4.3589	&4.5826 &5      &5.1962 &5.2915 \\
\textbf{N}	&1	&3	&6	&3	&6	&6	&6	&6	&3	&6	&12	&3	&6	&6	\\
\hline
\textbf{d}	&5.5678 &6      &6.0828 &6.245  &6.5574 &6.9282	&7	&7.2111 &7.5498 &7.8102 &7.9373 &8      &8.1854 &8.544  \\
\textbf{N}	&6	&6	&6	&12	&6	&6	&9	&6	&12	&6	&12	&3	&6	&6	\\
\hline
\textbf{d}	&8.6603 &8.7178	&8.8882 &9	&9.1652	&9.5394 &9.6437 &9.8489 &10     &10.1489&10.3923&10.4403&10.5357&	\\
\textbf{N}	&6	&6	&6	&6	&12	&12	&12	&6	&3	&6	&6	&6	&12	& \\
\hline
\end{tabular}
\end{table}

\begin{table}[h]
\caption{Distance between an atom on one graphene plane and nearest neighbors on next plane, for type II of neighbors, as in Fig. \ref{ringsAAAB} (right), found in AB stacking of layers. \textbf{N} is number of neighbors and \textbf{d} is distance (in units of graphene unit cell value of 1.42  {\AA}).}
\label{table_AB}
\centering
\begin{tabular}{|c|ccccccccccccc|}
\hline
\textbf{d}	&1	&2	&2.6458	&3.6056	&4	&4.3589	&5	&5.2915	&5.5678	&6.0828	&6.5574	&7 &7.2111\\
\textbf{N}	&6	&6	&12	&12	&6	&12	&6	&12	&12	&12	&12	&18 &	12\\
\hline
\textbf{d}	&7.8102	&8	&8.1854	&8.544	&8.7178	&8.8882	&9.5394	&9.8489	&10	&10.1489&10.4403&10.583	&\\
\textbf{N}	&12		&6	&12	&12	&12	&12	&24	&12	&6	&12	&12	&12	&\\
\hline
\end{tabular}
\end{table}

\end{widetext}

Potential energy $\Phi _{I}(z)$ for an atom in position AA at a distance $z$ from the plane will be given by an infinite sum of contributions from equi-distant atoms of type $I$:

\begin{equation}\label{MN_02}
\Phi _{I}(z) = \sum _{i=1}^{\infty} N(i) \cdot U\left(\sqrt{(z^2 + (a \cdot d_i)^2})\right),
\end{equation}

where the number of neighbors $N(i)$ at distance $d(i)$ is taken from entries of Table \ref{table_AA} and $U$ is any of potentials discussed (LJ, Mie, KC or Lebedeva's). In Eq. \ref{MN_02}, $a$ is the unit cell length for in-plane atoms ($a=1.42$ {\AA} in case of graphene).

In case of AB planes, there is equal number of atoms that are in ordering of type $I$ and type $II$. Equation on energy for type $II$ atoms, $\Phi _{II}(z)$, is similar as \ref{MN_02}, with summation taken on the data in Table \ref{table_AB}. We must take an average of $\Phi _{I}(z)$ and $\Phi _{II}(z)$ as an average energy of each atom in case of AB ordering. 

\begin{figure}[ht]
\centering
\includegraphics[scale=0.7]{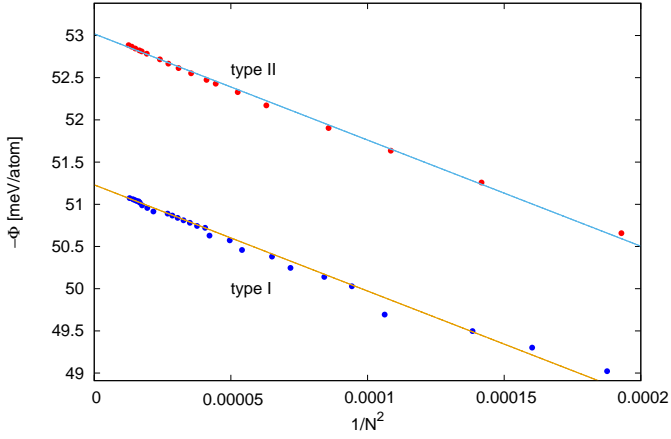}
      \caption{\label{DFT-fitting03b}Scaling of potential energy depth as a function of $1/N^2$, the number of atoms in a symmetric "molecule" for type I and II molecules, computed for LJ potential with parameters $\epsilon=2.80~meV$ and $\sigma=3.38~ {\AA}$. Asymptotic splitting for $N~\rightarrow~\infty$ between type I and type II energy minima, $(\Phi_{II}-\Phi_{I})/\Phi_{II}$, is $0.034$. Potential of type I is related to potential in AA stacking, while potential in AB stacking is an average of potentials of type I and II.
}
\end{figure}

Figure \ref{DFT-fitting03b} illustrates quick convergence of potential energy $\Phi$ as a function of the total number of atoms $N$ in a symmetric "molecule" for type I and type II "molecules" computed for LJ potential. An approximately $1/N^2$ scaling, when the rings of equi-distant neighbours are added in Eq. \ref{MN_02}.

Hence, an average potential energy for an atom at distance $z$ from the surface of bulk graphite is given as a sum, for AA, AB, and ABC stackings, respectively:

\begin{equation}\label{MN_04}
E_{AA}(z) = \sum _{i=1}^{\infty} \Phi _{I}(z + ic).
\end{equation}


\begin{widetext}

\begin{equation}\label{MN_03}
E_{AB}(z) = \sum _{i=1}^{\infty} \frac{1}{2} \left(\Phi _{I}(z + (2i-1)c) + \Phi _{II}(z + (2i-1)c)\right)+ \sum _{i=1}^{\infty} \Phi _{I}(z + 2ic),
\end{equation}

\begin{align}
\label{MN_05}
E_{ABC}(z) & = \frac{1}{2} \sum _{i=1}^{\infty} \left[ \Phi _{I}(z + 3ic)+\Phi _{II}(z + 3ic)\right]\nonumber\\
           & + \left[\Phi _{I}(x + (3i-1)c) + \Phi _{II}(x + (3i-1)c)\right] \nonumber\\
	   & + \left[\Phi _{I}(z + (3i-2)c)+\Phi _{II}(z + (3i-2)c)\right].
\end{align}
\end{widetext}

\begin{figure}[ht]
\centering
\includegraphics[scale=0.7]{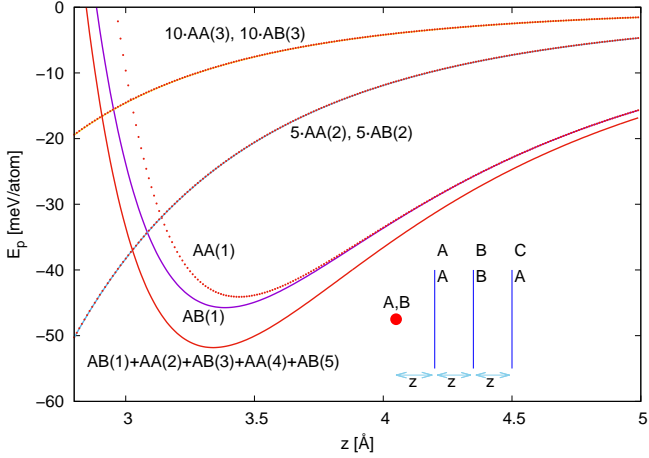}
      \caption{\label{ABC00a}Contribution to potential energy of a single atom (represented by large red dot in the insert) as a function of distance $z$ from the surface of three neighbouring planes, separated also by distance $z$. The case of ABA and ABC ordering of planes is considered. The atom is considered in two positions with respect to the first plane, AA(1) or AB(1). The energy contribution from the third plane is exactly the same regardless whether it is A plane or C plane. Energy from the second and third planes are rescaled 5 and 10 times, respectively. Classical L-J potential 12-6 with parameters $\epsilon=2.80~meV$ and $\sigma=3.38~ {\AA}$ was used. The depth of potential well for the lowest curve denoted as "AB(1)+AA(2)+AB(3)+AA(4)+AB(5)" (which shows the sum of energies from 5 consecutive planes in AB configuration) is 52 meV. 
}
\end{figure}

In our numerical computation we limit the number of planes in Eqs \ref{MN_04}--\ref{MN_05} to 4, since contribution to potential energy from next planes diminishes quickly: it is of the order of 90\%, 10\%, 2\% and 0.5\% for the first, second, third and forth plane, respectively (Fig. \ref{ABC00a}). Considering 4 planes only is equivalent to assuming cut-off distance of $5c$, that is 16.7 {\AA}.

The proposed method of computing interlayer potential is convenient for quick testing of potentials other than that of Mie or Lennard-Jones as well, since that requires only replacement of the function $U(z)$ in Eq. \ref{MN_02} by another one.

In case of perfect AA or AB stacking, due to symmetry, $\pi$-orbitals must point out in directions normal to planes.
In that case, in Kolmogorov-Crespi Equations \ref{eq_guerra}, values of ${\bf n}_l {\bf r}$ and ${\bf n}_m {\bf r}$ reduce to $z$, where $z$ is distance of an atom from the plane. That is, $\rho _{lm}$ and $\rho _{ml}$ become equal to $d$ from Tables \ref{table_AA} and \ref{table_AB}. We may rewrite Equations \ref{eq_guerra} in this form:

\begin{widetext}
\begin{equation}
U_{KC}(x) = -A\left(\frac{z_0}{r}\right)^6 + exp \left(-\lambda(r-z_0) \right) \cdot \left[C + 2exp\left( -(d/\delta )^2\right) \cdot \sum_{n=0}^{2} C_{2n} \left( d/\delta \right) ^{2n} \right],
\label{eq_guerra_simple}
\end{equation}


Similarly, we may rewrite Lebedeva's version of potential:

\begin{align}
U_{Leb}(x) = -A\left(\frac{z_0}{r}\right)^6 + B exp \left(-\alpha(r-z_0) \right)
  + C \left(1+D_1 d^2 + D_2 d^4  \right) exp \left( - \lambda _1 d ^2 \right) exp \left( - \lambda _2 \left( z^2 - z_0^2\right) \right).
\label{eq_lebedeva_simple}
\end{align}
\end{widetext}

In \ref{eq_guerra_simple} and \ref{eq_lebedeva_simple}, $r=\sqrt{z^2+d^2}$. 

\begin{widetext}

\begin{table}[h]
\caption{Summary of parameters' values used in Kolmogorov-Crespi equation \ref{eq_guerra_simple}.}
\label{table_KC_parameters}
\centering
\begin{tabular}{|ccccccccc|}
\hline
A [meV]	&C [meV]&$C_0 [meV]$	&$C_2 [meV]$	&$C_4 [meV]$	&$z_0 $[{\AA}]	&$\lambda$ [{\AA}$^{-1}$]	&$\delta$ [{\AA}]	&Ref.\\
10.238	& 3.030	& 15.71		& 12.29		& 4.933		& 3.34		& 3.629			& 0.578			&\cite{Kolmogorov2},\cite{Guerra}\\
9.89		& 0.7		&5.255    		& 2.336    		& 1.168		& 3.449		& 3.2				& 2.1				& this work, set A\\
8.881		& 0.6286	& 4.719    		& 2.098    		& 1.049		& 3.487		& 3.4				& 1.7				& this work, set B\\
\hline
\end{tabular}
\end{table}

\begin{table}[h]
\caption{Summary of parameters' values used in Lebedeva equation \ref{eq_lebedeva_simple}.}
\label{table_LEB_parameters}
\centering
\begin{tabular}{|cccccccccc|}
\hline
A [meV]	&B [meV]&C [meV]	&$\alpha [{\AA}^{-1}]$	&$z_0$ [{\AA}]	&$\lambda _1 [{\AA}^{-2}]$	&$\lambda _2$ [{\AA}$^{-2}]$	& $D_1$ [{\AA}$^{-2}]$	& $D_2$ [{\AA}$^{-4}]$	& Ref.\\
10.510	& 11.652	& 35.883	&  4.16		& 3.34		& 0.48703	& 0.46445	& -0.86232	& 0.10049	& \cite{Lebedeva}\\
14.558	& 21.204	& 1.8   	& 4.16    	& 3.198		&  0.6		& 0.4			& -0.862		& 0.10049	& this work\\
\hline
\end{tabular}
\end{table}
\end{widetext}

In case of Lennard-Jones potential, results presented here are computed with $\epsilon=2.4 meV$ and $\sigma=3.4322$. For models of Kolmogorov-Crespi and Lebedeva et al. we used values of parameters as listed in tables \ref{table_KC_parameters} and \ref{table_LEB_parameters}, respectively\footnote{LAMMPS implementation of equation of Lebedeva et al. is available from the corresponding author.}. One ought to be aware that usually there is a broad range of parameters values for any of the above models of potential that may provide a functionally nearly identical dependencies $U(z)$ and reasonable agreement with experiments; compare for instance \cite{Kolmogorov2},\cite{Guerra} with \cite{Kolmogorov} and \cite{Annelot}. The original parameters of Lebedeva equation were derived by fitting to results of $ab-initio$ DFT modeling and with the purpose of describing graphene corrugation experiments. While we recognize the usefulness of proposed new equations for providing phenomenological description of certain materials properties, we find no strong justification for adhering to original values of  parameters. In particular, as we discuss later, the value of $C$ in Table \ref{table_LEB_parameters} proposed in the original work is likely significantly too large.

The meaning of some parameters is as follows. $\sigma$ and $z_0$ decide most about position of potential minimum (equilibrium interlayer spacing), which is also sensitive to $D_2$ and $\lambda _2$, while $\lambda$, $\delta$, $\lambda _1$ and mainly $C$ in Eq.  \ref{eq_lebedeva_simple} decide most about AB-AA energy difference at potentials minima. Value of $\lambda$ influences mostly the slope of $dE/dz$ in Fig. \ref{ABC00a} on low-z side of potential curve and as a consequence it decides about the compressibility as a function of pressure. 

\subsection{Compressibility.}

Equation \ref{MN_03} may be used for computing compressibility along c-axis, since force and pressure acting on a particle 
is proportional to derivative of potential energy:

\begin{equation}\label{compressibility}
P(c) = \eta \cdot \frac{dE_{AB}(c)}{dc},
\end{equation}

where $c$ is lattice constant perpendicular to the planes under pressure $P$ and $\eta$ is the number of atoms per unit area, $\eta=61.171~GPa \cdot eV^{-1} \cdot {\AA}^{-3}$ for graphene. The first (Eq. \ref{compressibility}) and second derivatives of $E_{AB}(z)$ could be computed by using derived analytical expressions. Numerical approach of finding derivatives is however convenient to implement and fast. 

\begin{figure}[ht]
\centering
\includegraphics[scale=0.7]{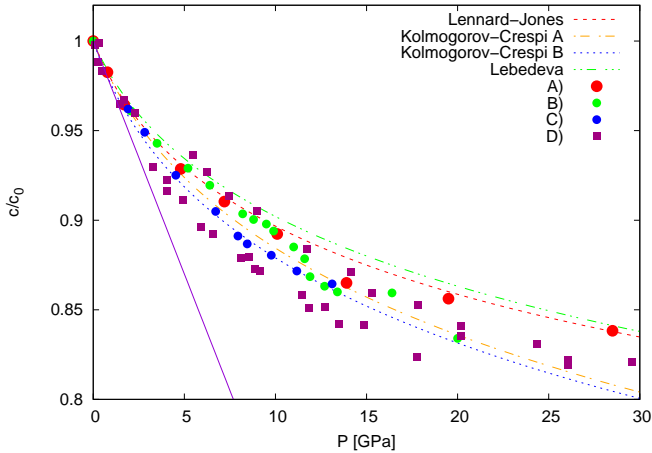}
      \caption{\label{compress02a}Compressibility of graphite along c-axis. 
The solid curves are computed numerically by using Eq. \ref{compressibility} with L-J, K-C, and Lebedeva potentials. 
Datapoints are collected from works of Lynch and Drickamer (\cite{Lynch}, A), Zhao and Spain (\cite{ZhaoSpain}, B), Hanfland et al. (\cite{Hanfland}, C), Clark et al. (\cite{Clark}, D). 
}
\end{figure}

Figure \ref{compress02a} shows the found ratio of $c/c_0$ as a function of pressure for LJ, KC and Lebedeva potentials, with parameters as these in Tables \ref{table_KC_parameters} and \ref{table_LEB_parameters}. In case of LJ model, the parameter $\epsilon$ used was $3.3 meV$, which leads to binding energy of $64 meV/atom$, larger than most commonly accepted value of around $50 meV/atom$ but leading to better agreement between results of other calculations and measurements, as reported in the following sections. The straight solid line with a slope of $-0.029/GPa$ at $P=0$ for LJ potential corresponds to elasticity modulus $C_{33}=38.5~GPa$, and for KC one the slope gives $C_{33}=43.5~GPa$, in a very good agreement with experiment. $Ab-initio$ DFT computations result in a broader spread of values, from $43.6$ to $67.5~GPa$ \cite{Gao}. 

Pressure dependence of $c$ can be found or deduced from several measurements (\cite{Wang}, \cite{Lynch}, \cite{ZhaoSpain}, \cite{Clark}, \cite{Trucano}). Experimental results in Fig. \ref{compress02a} have a broad dispersion of data points, which is, in part, due to subtle differences in the used measurement techniques.

\subsection{Stacking order}

There is much controversy around prevailing stacking order in graphite and in a few layers graphene (FLG).

It is known that there are three types of stacking, an AA one, which is simple hexagonal (it is not found in natural graphite and exists only in intercalated compounds such as $C_6Li$ and $C_8K$, 
an AB stacking, hexagonal, known also as Bernal, and a rhombohedral ABC stacking. Additionally, a random stacking of these three types is called a turbostratic TS structure and often is obtained in laboratories. In bulk graphite, it was reported that the volume fraction AB:ABC:TS was about 80:14:6 (\cite{Lipson}).

Lee et al. \cite{LeeLee} were able to obtain AA structured graphene on diamond surface, with an interplanar spacing of $\sim $3.55{\AA}. That value is between that of the AB graphite ($3.35${\AA}) and the Lithium intercalated AA graphite $3.706${\AA} \cite{Dahn}.
Norimatsu and Kusunoki \cite{Norimatsu} observe ABC-stacked graphene on the $SiC$ substrate and interpret their results in terms of possible modification of second-plane interactions by the substrate, as explained by Slonczewski-Weiss-McClure model \cite{Charlier3}. 
Yoshizawa et al. \cite{Yoshizawa} propose that the AB stacking of layers in graphite is a consequence of orbital interactions between layers rather than of generally accepted van der Waals forces.

The energy difference between AA and AB is reported, to be 17.31 meV/atom (\cite{Mostaani}), in favour of AB, while between ABC and AB it was reported as 0.11 meV/atom (\cite{Charlier}), which are both based however on DFT calculations, only.

It was observed that in the case of graphene, its bilayer exhibits AB stacking while trilayer prefers ABC stacking \cite{WangGao}. When the number of layers increases, again AB stacking is favored. The binding energies are found to increase from $23 meV/atom$ in bilayer to $39 meV/atom$ in pentalayer, while the interlayer distance decreases from $3.37 {\AA}$ in bilayer to $3.35  {\AA}$ in pentalayer. Our models do not reproduce so strong change of binding energy with number of layers. In 2-layer graphene we find around 90\% of binding energy per atom of that in bulk graphite, when LJ model is used  (Fig. \ref{ABC00a}). However, the change of interlayer distance agrees with that reported \cite{Tao}, for all three models discussed.

\begin{widetext}

\begin{table}[h]
\caption{Energies of FLG with number of layers from 2 to 6, for AA, AB and ABC stacking.}
\label{table_ABC}
\centering
\begin{tabular}{|c|ccccc|}
\hline
ordering 	&2 layers		&3 layers		&4 layers				&5 layers					&6 layers\\
\hline
AA	&AA(1)		&2AA(1)+AA(2)	&3AA(1)+2AA(2)+AA(3)		&4AA(1)+3AA(2)+2AA(3)		&5AA(1)+3AA(2)+3AA(3)+2AA(4)\\
AB	&AB(1)		&2AB(1)+AA(2)	&3AB(1)+2AA(2)+AB(3)		&4AB(1)+3AA(2)+2AB(3)		&5AB(1)+4AA(2)+3AB(3)+2AA(4)\\
ABC	&--			&2AB(1)+AA(2)	&3AB(1)+2AB(2)+AA(3)		&4AB(1)+3AB(2)+2AA(3)		&5AB(1)+4AB(2)+3AA(3)+2AB(4)\\
\hline
\end{tabular}
\end{table}

\end{widetext}

If to assume that interaction between next nearest planes (NNP) is negligible than there is no reason for difference in stability of ABC and AB structures. Hence (NNP) interaction must play a role. Table \ref{table_ABC} summarizes total potential energy of a FLG. The notation used there is the same as in Fig. \ref{ABC00a}, and it should be understood as follows (taking AB stacking for 3 layers, as an example):  2AB(1)+AA(2) means that we have 2 pairs of layers in AB stacking that are apart for $1c$, and one pair of AA ordering of layers apart for $2c$. One should notice that energy of AC stacking is the same as that of AB, since atoms configuration of neighboring planes are the same in both cases. We see from Table \ref{table_ABC} that for 4 and more layers ABC stacking is energetically favorable. However the energy difference between AB and ABC is caused by interaction of layers that are $2c$ apart and therefore it is very small, of around $3 \mu eV/atom$.

The explanation to AB:ABC ratio observed, of about 80:14 (\cite{Lipson}), could be found by adding to already existing isotropic attractive term of vdW potential
a small anisotropic contribution to equations like these of KC or Lebedeva.  Since interplane binding is caused by strongly anisotropic interaction between $\pi$ orbitals one would expect that not only repulsive but also attractive component of that interaction is anisotropic.

\subsection{Raman shifts in LBM and shear modes ($C_{33}$ and $C_{44}$ elastic constants)}

Early neutron scattering measurements confirmed the existence of the low-frequency acoustic  mode in
bulk pyrolitic graphite \cite{Dolling}, that is longitudinal waves in the direction of the hexagonal axis, at $3.84\pm 0.06~THz$. The transverse waves (in shear mode) were less pronounced, at $1.3 \pm 0.3~THz$. Based on these data elastic constants have being deduced, $C_{33}=39\pm 4 GPa$ (for direction perpendicular to graphite planes; LBM), and $C_{44}=4.2\pm 2 GPa$ (shear mode parallel to planes). These values correspond to Raman shift energies of 130 and 43 $cm^{-1}$, in agreement with several DFT calculations \cite{Lebedeva4}. The experimental evidence of the longitudinal mode was reported in ultrafast laser pump-probe spectroscopy \cite{Shang} on FLG, at $\sim 120 cm^{-1}$, demonstrating also the presence  there of the phonon and electron coupling through the Breit-Wigner-Fano resonance, as at more pronounced G-mode \cite{Baranowski}.

Position of peaks in both LBM and shear modes is a geometrical effect and depends strongly on the number of layers in FLG. Moreover, 
unique multipeak features are observed, characteristic for the number of layers investigated.
Their frequencies  in case of LBM \cite{Lui} mode  are described well using a simple linear-chain model based on nearest-neighbor couplings between the layers \cite{Thornton}:

\begin{equation}\label{LBM}
\omega_N(n) = \omega_0 \sin {[(N-n)\pi / 2N]},
\end{equation}

where $\omega_0$ is the frequency of the bulk mode, $N$ it the number of layers in FLG and $n$ enumerates observed Raman frequencies. It follows from
 measurements that in the limit of large number of layers the bulk LBM Raman peak should have value of $264.5~cm^{-1}$ \cite{Lui}. 

For shear mode, the nature of interactions between planes, their collective behaviour, leads to quantitatively similar dependence of Raman frequencies on the number of layers in FLG as in the case of LBM mode: the frequency in the bulk is $\sqrt{2}$ times larger than for bilayer graphene \cite{Tan}, \cite{Cong}.  Its frequency depends on energy difference between AB and AA stacking. For that reason it is sensitive to the choice of inter-plane interaction potential.

\begin{figure}[ht]
\centering
\includegraphics[scale=0.7]{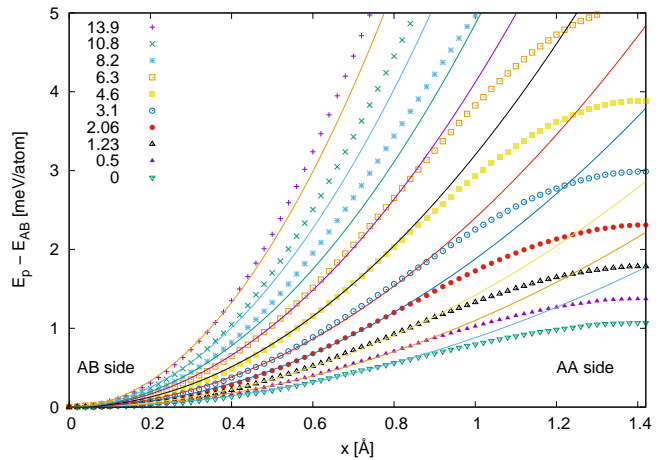}
      \caption{\label{Mie_Ep8C_ABAA_plane_fit01}Datapoints show change in potential energy of a plane that is moved from AB position over
another large plane, for distance equal to a=1.42 {\AA}, which results in AA position, under several values of pressure [GPa], as listed in the legend. 
The classical 12-6 Lennard-Jones potential was used in modelling, with $\sigma=3.4322 {\AA}$ and $\epsilon=2.4 meV$.
Solid lines show a tentative fit of parabolic dependencies.
}
\end{figure}

In Fig. \ref{Mie_Ep8C_ABAA_plane_fit01} we show how the potential energy changes when a graphene plane lying in a distance of 3.35 {\AA} over another graphene plane moves away from AB position for 1.42 {\AA}, towards AA stacking. These calculations were made in LAMMPS, assuming LJ potential with $\sigma=3.4322 {\AA}$ and $\epsilon=2.4 meV$. We find that the energy difference between planes in AA and AB positions is of around $1 meV/atom$, only, at $P=0~GPa$.

By having the curvature of parabolic potential wells at low-values of departure from optimal position of both layers (in AB stacking; we used data for x less then 0.05 {\AA})  we are able to compute Raman frequencies in shear mode in a broad range of pressure. The curvature $k$ of parabolic fit of data in Figure \ref{Mie_Ep8C_ABAA_plane_fit01}, $E_p(x)-E_{AB}=k/2\cdot x^2$, with equation $R[cm^{-1}]=\sqrt{2}\cdot \sqrt{k} \cdot 703$, where $k$ is in $eV/{\AA}^2$ and $R$ is the Raman shift. The $\sqrt{2}$ factor is to account that the bulk potential acting on a plane is twice as large as potential for single plane on the surface of sample. 

\begin{figure}[ht]
\centering
\includegraphics[scale=0.7]{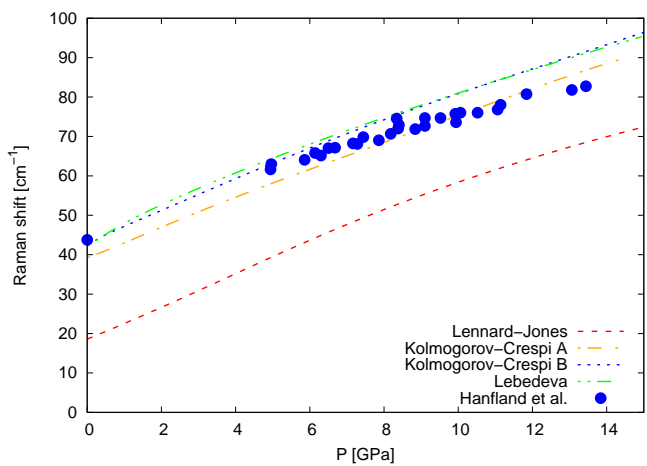}
      \caption{\label{raman00}
Raman shift energy as a function of pressure, for bulk shear mode. The lines are computed from curvature $k$ of parabolic fit of data in Figure \ref{Mie_Ep8C_ABAA_plane_fit01}.  Blue circles are experimental data of Hanfland et al. \cite{Hanfland}.
}
\end{figure}

Results on Raman shift are shown in Fig. \ref{raman00}, where a comparison is made of the data obtained by using LJ, Lebedeva and KC potential models (for KC with two sets of parameters), as listed in Tables \ref{table_KC_parameters}, \ref{table_LEB_parameters}. The data in Fig. \ref{raman00} are compared with experimental results of Hanfland et al. \cite{Hanfland}. It is evident that LJ potential can not describe well measured values of Raman shift while both, KC and Lebedeva models, allow to achieve an acceptable fit to real data. We observe also that LJ frequency shift at $P=0 GPa$ is about twice too low to explain the Raman shift observed in experiment . This means that the energy difference between AA and AB arrangement of planes must be around 5 meV per atom at $P=0~GPa$, which is significantly smaller than reported in some DFT calculations \cite{Mostaani}, \cite{Lebedeva}.

For LBM mode results on the Raman shift are known at $P=0 GPa$, only, and our model predicts that value accurately \cite{Lui}. In this case there is no need to use LAMMPS in computation. The scheme described in this section offers an alternate and convenient way of numerical computing frequency of oscillations from potential energy curvature by starting with Eq. \ref{MN_03}. We need to use also the data available in Fig. \ref{compress02a} and find out from there how planes' spacing $c$ changes with pressure. We insert $c$ values into Eq. \ref{MN_03} and numerically find out the second derivative $d^2E_{AB}(z)/dz^2$, which gives us curvature of potential minimum and, next, frequencies of Raman shift, as a function of pressure.

\begin{figure}[ht]
\centering
\includegraphics[scale=0.7]{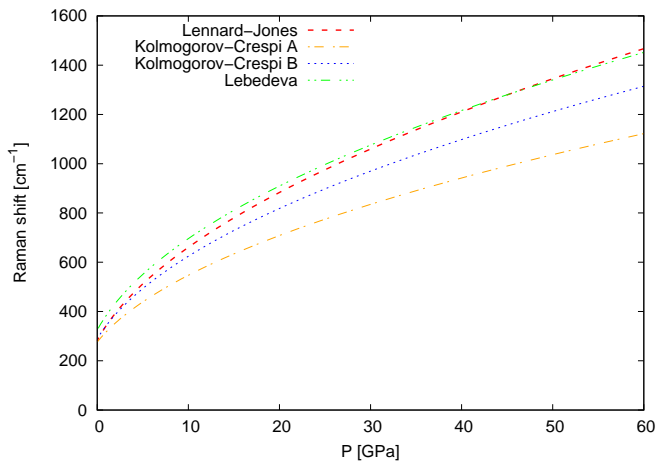}
      \caption{\label{ABC_LBM}Predicted pressure dependence of Raman shift for LBM mode of bulk graphite.
}
\end{figure}

Figure \ref{ABC_LBM} shows pressure dependence of Raman shift in LBM mode computed for LJ, KC and Lebedeva models, by using this method. If similar experimental data were available we could have an additional indication on which of considered models of vdW interactions fits best to reality. 

%% file: graphene.bbl
\begin{thebibliography}{76}
\providecommand{\natexlab}[1]{#1}
\providecommand{\url}[1]{\texttt{#1}}
\expandafter\ifx\csname urlstyle\endcsname\relax
  \providecommand{\doi}[1]{doi: #1}\else
  \providecommand{\doi}{doi: \begingroup \urlstyle{rm}\Url}\fi

\bibitem[Geim and Grigorieva(2013)]{Geim}
A.~K. Geim and I.~V. Grigorieva.
\newblock Van der waals heterostructures.
\newblock \emph{Nature}, 499:\penalty0 419--425, 2013.
\newblock \doi{10.1038/nature12385}.
\newblock URL \url{http://dx.doi.org/10.1038/nature12385}.

\bibitem[Le et~al.(2016)Le, Huan, and Woods]{Nam}
Nam~B. Le, Tran~Doan Huan, and Lilia~M. Woods.
\newblock Interlayer interactions in van der waals heterostructures: Electron
  and phonon properties.
\newblock \emph{ACS Appl. Mater. Interfaces}, 9(9):\penalty0 6286--6292, 2016.
\newblock \doi{10.1021/acsami.6b00285}.
\newblock URL \url{http://dx.doi.org/10.1021/acsami.6b00285}.

\bibitem[van Wijk et~al.(2015)van Wijk, Schuring, Katsnelson, and
  Fasolino]{Wijk}
M~M van Wijk, A~Schuring, M~I Katsnelson, and A~Fasolino.
\newblock Relaxation of moiré patterns for slightly misaligned identical
  lattices: graphene on graphite.
\newblock \emph{2D Materials}, 2(3):\penalty0 034010, 2015.
\newblock \doi{10.1088/2053-1583/2/3/034010}.
\newblock URL \url{http://doi.org/10.1088/2053-1583/2/3/034010}.

\bibitem[van Wijk et~al.(2014{\natexlab{a}})van Wijk, Schuring, Katsnelson, and
  Fasolino]{Wijk2}
M~M van Wijk, A~Schuring, M~I Katsnelson, and A~Fasolino.
\newblock Moiré patterns as a probe of interplanar interactions for graphene
  on h-bn.
\newblock \emph{Phys. Rev. Lett.}, 113:\penalty0 135504, 2014{\natexlab{a}}.
\newblock \doi{10.1103/PhysRevLett.113.135504}.
\newblock URL \url{http://doi.org/10.1103/PhysRevLett.113.135504}.

\bibitem[Cao et~al.(2018)Cao, Fatemi, Fang, Watanabe, Taniguchi, Kaxiras, and
  Jarillo-Herrero]{YuanCao}
Yuan Cao, Valla Fatemi, Shiang Fang, Kenji Watanabe, Takashi Taniguchi,
  Efthimios Kaxiras, and Pablo Jarillo-Herrero.
\newblock Unconventional superconductivity in magic-angle graphene
  superlattices.
\newblock \emph{Nature}, 2018.
\newblock \doi{http://dx.doi.org/10.1038/nature26160}.
\newblock URL \url{http://dx.doi.org/10.1038/nature26160}.

\bibitem[Peng et~al.(2017)Peng, Schröter, Yin, Wang, Chung, Yang, Ekahana,
  Liu, Jiang, Yang, Zhang, Cheng~Chen, Barinov, Chen, Liu, Peng, and
  Chen]{HanPeng}
Han Peng, Niels B.~M. Schröter, Jianbo Yin, Huan Wang, Ting-Fung Chung,
  Haifeng Yang, Sandy Ekahana, Zhongkai Liu, Juan Jiang, Lexian Yang, Teng
  Zhang, Heng~Ni Cheng~Chen, Alexey Barinov, Yong~P. Chen, Zhongfan Liu, Hailin
  Peng, and Yulin Chen.
\newblock Substrate doping effect and unusually large angle van hove
  singularity evolution in twisted bi- and multilayer graphene.
\newblock \emph{Adv. Mater.}, page 1606741, 2017.
\newblock \doi{10.1002/adma.201606741}.
\newblock URL \url{http://doi.org/10.1002/adma.201606741}.

\bibitem[Argentero et~al.(2017)Argentero, Mittelberger, Monazam, Cao,
  Pennycook, Mangler, Kramberger, Kotakoski, Geim, , and Meyer]{Argentero}
Giacomo Argentero, Andreas Mittelberger, Mohammad Reza~Ahmadpour Monazam, Yang
  Cao, Timothy~J. Pennycook, Clemens Mangler, Christian Kramberger, Jani
  Kotakoski, A.~K. Geim, , and Jannik~C. Meyer.
\newblock Unraveling the 3d atomic structure of a suspended graphene/hbn van
  der waals heterostructure.
\newblock \emph{Nano Lett.}, 17:\penalty0 1409--1406, 2017.
\newblock \doi{http://doi.org/10.1021/acs.nanolett.6b04360}.
\newblock URL \url{http://doi.org/10.1021/acs.nanolett.6b04360}.

\bibitem[Lin et~al.(2017)Lin, Lu, Xu, Feng, and Li]{Lin}
Shisheng Lin, Yanghua Lu, Juan Xu, Sirui Feng, and Jianfeng Li.
\newblock High performance graphene/semiconductor van der waals heterostructure
  optoelectronic devices.
\newblock \emph{Nano Energy}, 40:\penalty0 122--148, 2017.
\newblock \doi{10.1016/j.nanoen.2017.07.036}.
\newblock URL \url{https://doi.org/10.1016/j.nanoen.2017.07.036}.

\bibitem[Skoblin et~al.(2018)Skoblin, Sun, and Yurgens]{Skoblin}
Grigory Skoblin, Jie Sun, and August Yurgens.
\newblock Graphene bolometer with thermoelectric readout and capacitive
  coupling to an antenna.
\newblock \emph{Appl. Phys. Lett.}, 112:\penalty0 063501, 2018.
\newblock \doi{http://doi.org/10.1063/1.5009629}.
\newblock URL \url{https://doi.org/10.1063/1.5009629}.

\bibitem[Wei et~al.(2017)Wei, Yan, Shen, Lv, and Wang]{Wei}
Xia Wei, Fa-Guang Yan, Chao Shen, Quan-Shan Lv, and Kai-You Wang.
\newblock Photodetectors based on junctions of two-dimensional transition metal
  dichalcogenides.
\newblock \emph{Chinese Phys. B 26}, 26 (3):\penalty0 038504, 2017.
\newblock \doi{10.1088/1674-1056/26/3/038504}.
\newblock URL \url{https://doi.org/10.1088/1674-1056/26/3/038504}.

\bibitem[Congpu et~al.(2017)Congpu, Jianyong, and Zhongyuan]{muxiang}
Mu~Congpu, Xiang Jianyong, and Liu Zhongyuan.
\newblock Photodetectors based on sensitized two-dimensional transition metal
  dichalcogenides—a review.
\newblock \emph{Journal of Materials Research}, 32\penalty0 (22):\penalty0
  4115–4131, 2017.
\newblock \doi{10.1557/jmr.2017.402}.

\bibitem[Houssa et~al.(2015)Houssa, Dimoulas, and Molle]{Houssa}
M~Houssa, A~Dimoulas, and A~Molle.
\newblock Silicene: a review of recent experimental and theoretical
  investigations.
\newblock \emph{Journal of Physics: Condensed Matter}, 27(25):\penalty0 253002,
  2015.
\newblock URL \url{http://stacks.iop.org/0953-8984/27/i=25/a=253002}.

\bibitem[Kvashnin and Chernozatonskii(2017)]{Kvashnin}
D.~G. Kvashnin and L.~A. Chernozatonskii.
\newblock Electronic and transport properties of heterophase compounds based on
  mos2.
\newblock \emph{Jetp Lett.}, 105:\penalty0 250, 2017.
\newblock \doi{10.1134/S0021364017040117}.
\newblock URL \url{http://doi.org/10.1134/S0021364017040117}.

\bibitem[Luo et~al.(2017{\natexlab{a}})Luo, Shen, and Yin]{MLuo}
M.~Luo, Y.~H. Shen, and T.~L. Yin.
\newblock Structural, electronic, and magnetic properties of transition metal
  doped res2 monolayer.
\newblock \emph{Jetp Lett.}, 105:\penalty0 255, 2017{\natexlab{a}}.
\newblock \doi{10.1134/S0021364017040038}.
\newblock URL \url{http://doi.org/10.1134/S0021364017040038}.

\bibitem[Bellus et~al.(2017)Bellus, Li, Lane, Ceballos, Cui, Zeng, and
  Zhao]{Bellus}
Matthew~Z. Bellus, Ming Li, Samuel~D. Lane, Frank Ceballos, Qiannan Cui,
  Xiao~Cheng Zeng, and Hui Zhao.
\newblock Type-i van der waals heterostructure formed by mos2 and res2
  monolayers.
\newblock \emph{Nanoscale Horiz.}, 2:\penalty0 31--36, 2017.
\newblock \doi{10.1039/C6NH00144K}.
\newblock URL \url{http://doi.org/10.1039/C6NH00144K}.

\bibitem[Yan et~al.(2017)Yan, Zhao, Patanè, Hu, Wei, Luo, Zhang, Lv, Feng,
  Shen, Chang, Eaves, and Wang]{FYan}
Faguang Yan, Lixia Zhao, Amalia Patanè, PingAn Hu, Xia Wei, Wengang Luo, Dong
  Zhang, Quanshan Lv, Qi~Feng, Chao Shen, Kai Chang, Laurence Eaves, and Kaiyou
  Wang.
\newblock Fast multicolor photodetectors based on graphene-contacted
  p-gase/n-inse van der waals heterostructures.
\newblock \emph{Nanotechnology}, 28\penalty0 (7):\penalty0 27LT01, 2017.
\newblock \doi{http://doi.org/10.1088/1361-6528/aa749e}.
\newblock URL \url{http://stacks.iop.org/0957-4484/28/i=27/a=27LT01}.

\bibitem[Luo et~al.(2017{\natexlab{b}})Luo, Yin, and Chu]{LuoYin}
M.~Luo, H.~H. Yin, and J.~H. Chu.
\newblock Magnetic properties of a na-doped ws2 monolayer in the presence of an
  isotropic strain.
\newblock \emph{JETP Letters}, 6\penalty0 (10):\penalty0 672--676,
  2017{\natexlab{b}}.
\newblock \doi{http://doi.org/10.1134/S0021364017220039}.
\newblock URL \url{http://doi.org/10.1134/S0021364017220039}.

\bibitem[Sanchez et~al.(2016)Sanchez, Ovchinnikov, Misra, Allain, and Kis]{OLS}
Oriol~Lopez Sanchez, Dmitry Ovchinnikov, Shikhar Misra, Adrien Allain, and
  Andras Kis.
\newblock Valley polarization by spin injection in a light-emitting van der
  waals heterojunction.
\newblock \emph{Nano Lett.}, 16:\penalty0 5792--5797, 2016.
\newblock \doi{http://doi.org/10.1021/acs.nanolett.6b02527}.

\bibitem[Gurram et~al.(2017)Gurram, Omar, and van Wees]{Gurram}
Mallikarjuna Gurram, Siddhartha Omar, and Bart~J. van Wees.
\newblock Bias induced up to 100
  ferromagnet/bilayer-hbn/graphene/hbn heterostructures.
\newblock \emph{Nature Communications}, 8:\penalty0 248, 2017.
\newblock \doi{http://doi.org/10.1038/s41467-017-00317-w}.

\bibitem[Shim et~al.(2017)Shim, Jo, Kim, Song, Kim, and Park]{Shim}
Jaewoo Shim, Seo-Hyeon Jo, Minwoo Kim, Young~Jae Song, Jeehwan Kim, and
  Jin-Hong Park.
\newblock Light-triggered ternary device and inverter based on heterojunction
  of van der waals materials.
\newblock \emph{ACS Nano}, 11\penalty0 (6):\penalty0 6319--6327, 2017.
\newblock \doi{http://doi.org/10.1021/acsnano.7b02635}.

\bibitem[Li et~al.(2017)Li, Wang, and Dai]{WeiLi}
Wei Li, Xinlian Wang, and Xianqi Dai.
\newblock Tunable schottky contacts in the antimonene/graphene van der waals
  heterostructures.
\newblock \emph{Solid State Communications}, 254:\penalty0 37--41, 2017.
\newblock \doi{10.1016/j.ssc.2017.02.008}.
\newblock URL \url{https://doi.org/10.1016/j.ssc.2017.02.008}.

\bibitem[Michel et~al.(2017)Michel, Çakır, Sevik, and Peeters]{Michel}
K.~H. Michel, D.~Çakır, C.~Sevik, and F.~M. Peeters.
\newblock Piezoelectricity in two-dimensional materials: Comparative study
  between lattice dynamics and ab initio calculations.
\newblock \emph{Phys. Rev. B}, 95:\penalty0 125415, 2017.
\newblock \doi{https://doi.org/10.1103/PhysRevB.95.125415}.

\bibitem[Kawai et~al.(2016)Kawai, Foster, Björkman, Nowakowska, Björk,
  Canova, Gade, Jung, and Meyer]{Kawai}
Shigeki Kawai, Adam~S. Foster, Torbjörn Björkman, Sylwia Nowakowska, Jonas
  Björk, Filippo~Federici Canova, Lutz~H. Gade, Thomas~A. Jung, and Ernst
  Meyer.
\newblock Van der waals interactions and the limits of isolated atom models at
  interfaces.
\newblock \emph{Nature Communications}, 7:11559, 2016.
\newblock \doi{10.1038/ncomms11559}.
\newblock URL \url{http://dx.doi.org/10.1038/ncomms11559}.

\bibitem[Rozhkov et~al.(2016)Rozhkov, Sboychakov, Rakhmanov, and Nori]{Rozhkov}
A.V. Rozhkov, A.O. Sboychakov, A.L. Rakhmanov, and Franco Nori.
\newblock Electronic properties of graphene-based bilayer systems.
\newblock \emph{Physics Reports}, 648:\penalty0 1--104, 2016.
\newblock \doi{10.1016/j.physrep.2016.07.003}.
\newblock URL \url{http://dx.doi.org/10.1016/j.physrep.2016.07.003}.

\bibitem[Girifalco and Hodak(2002)]{Girifalco2}
L.~A. Girifalco and Miroslav Hodak.
\newblock Van der waals binding energies in graphitic structures.
\newblock \emph{Phys. Rev. B}, 65:\penalty0 125404, 2002.
\newblock \doi{10.1103/PhysRevB.65.125404}.
\newblock URL \url{http://doi.org/10.1103/PhysRevB.65.125404}.

\bibitem[Charlier et~al.(1994{\natexlab{a}})Charlier, Gonze, and
  Michenaud]{Charlier2}
J.-C. Charlier, X.~Gonze, and J.-P. Michenaud.
\newblock Graphite interplanar bonding: Electronic delocalization and van der
  waals interaction.
\newblock \emph{Europhysics Letters}, 28:\penalty0 403, 1994{\natexlab{a}}.
\newblock URL \url{http://stacks.iop.org/0295-5075/28/i=6/a=005}.

\bibitem[Trickey et~al.(1992)Trickey, Müller-Plathe, Diercksen, , and
  Boettger]{Trickey}
S.~B. Trickey, F.~Müller-Plathe, G.~H.~F. Diercksen, , and J.~C. Boettger.
\newblock Interplanar binding and lattice relaxation in a graphite dilayer.
\newblock \emph{Physical Review B}, 46:\penalty0 4460, 1992.
\newblock \doi{10.1103/PhysRevB.45.4460}.
\newblock URL \url{http://doi.org/10.1103/PhysRevB.45.4460}.

\bibitem[Rydberg et~al.(2003)Rydberg, Dion, Jacobson, Schröder, Hyldgaard,
  Simak, Langreth, and Lundqvist]{Rydberg}
H.~Rydberg, M.~Dion, N.~Jacobson, E.~Schröder, P.~Hyldgaard, S.~I. Simak,
  D.~C. Langreth, and B.~I. Lundqvist.
\newblock Van der walls density functional for layered structures.
\newblock \emph{Physical Review Letters}, 91:\penalty0 126402, 2003.
\newblock \doi{10.1103/PhysRevLett.91.126402}.
\newblock URL \url{https://doi.org/10.1103/PhysRevLett.91.126402}.

\bibitem[DiVincenzo et~al.(1983)DiVincenzo, Mele, and Holzwarth]{DiVincenzo}
D.~P. DiVincenzo, E.~J. Mele, and N.~A.~W. Holzwarth.
\newblock Density-functional study of interplanar binding in graphite.
\newblock \emph{Physical Review B}, 27:\penalty0 2458, 1983.
\newblock \doi{10.1103/PhysRevB.27.2458}.
\newblock URL \url{https://doi.org/10.1103/PhysRevB.27.2458}.

\bibitem[Schabel and Martins(1992)]{Schabel}
Matthias~C. Schabel and José~Luís Martins.
\newblock Energetics of interplanar binding in graphite.
\newblock \emph{Physical Review B}, 46:\penalty0 7185, 1992.
\newblock \doi{10.1103/PhysRevB.46.7185}.
\newblock URL \url{https://doi.org/10.1103/PhysRevB.46.7185}.

\bibitem[Mostaani et~al.(2015)Mostaani, Drummond, and Fal'ko]{Mostaani}
E.~Mostaani, N.~D. Drummond, and V.~I. Fal'ko.
\newblock Quantum monte carlo calculation of the binding energy of bilayer
  graphene.
\newblock \emph{Phys. Rev. Lett.}, 115:\penalty0 115501, 2015.
\newblock \doi{10.1103/PhysRevLett.115.115501}.
\newblock URL \url{https://doi.org/10.1103/PhysRevLett.115.115501}.

\bibitem[Birowska et~al.(2011)Birowska, Milowska, and Majewski]{Birowska}
M.~Birowska, K.~Milowska, and J.A. Majewski.
\newblock Van der waals density functionals for graphene layers and graphite.
\newblock \emph{Acta Physica Polonica A}, 120\penalty0 (5):\penalty0 845, 2011.
\newblock URL \url{http://przyrbwn.icm.edu.pl/APP/PDF/116/a116z520.pdf}.

\bibitem[Chakarova-Kack et~al.(2006)Chakarova-Kack, Schroder, Lundqvistand, and
  Langreth]{Chakarova}
S.~D. Chakarova-Kack, E.~Schroder, B.~I. Lundqvistand, and D.~C. Langreth.
\newblock Application of van der waals density functional to an extended
  system: Adsorption of benzene and naphthalene on graphite.
\newblock \emph{Phys. Rev. Lett.}, 96:\penalty0 146107, 2006.
\newblock \doi{10.1103/PhysRevLett.96.146107}.
\newblock URL \url{http://doi.org/10.1103/PhysRevLett.96.146107}.

\bibitem[Lu et~al.(1992)Lu, Li, , and Martin]{Ping}
Jian~Ping Lu, X.-P. Li, , and Richard~M. Martin.
\newblock Ground state and phase transitions in solid c60.
\newblock \emph{Phys. Rev. Lett.}, 68:\penalty0 1551, 1992.
\newblock \doi{10.1103/PhysRevLett.68.1551}.
\newblock URL \url{https://doi.org/10.1103/PhysRevLett.68.1551}.

\bibitem[Jiang et~al.(2006)Jiang, Huang, Jiang, Ravichandran, Hwang, and
  Liu]{Jiang}
L.Y. Jiang, Y.~Huang, H.~Jiang, G.~Ravichandran, H.~GaoandK.C. Hwang, and
  B.~Liu.
\newblock A cohesive law for carbon nanotube/polymer interfaces based on the
  van der waals force.
\newblock \emph{Journal of the Mechanics and Physics of Solids}, 54:\penalty0
  2436–2452, 2006.
\newblock \doi{10.1016/j.jmps.2006.04.009}.
\newblock URL \url{http://doi.org/10.1016/j.jmps.2006.04.009}.

\bibitem[He et~al.(2005)He, Kitipornchai, and Liew]{XQHe}
X.Q. He, S.~Kitipornchai, and K.M. Liew.
\newblock Buckling analysis of multi-walled carbon nanotubes: a continuum model
  accounting for van der waals interaction.
\newblock \emph{Journal of the Mechanics and Physics of Solids}, 53:\penalty0
  303--326, 2005.
\newblock \doi{10.1016/j.jmps.2004.08.003}.
\newblock URL \url{http://doi.org/10.1016/j.jmps.2004.08.003}.

\bibitem[Kitipornchai et~al.(2005)Kitipornchai, He, and Liew]{Kitipornchai}
S.~Kitipornchai, X.~Q. He, and K.~M. Liew.
\newblock Continuum model for the vibration of multilayered graphene sheets.
\newblock \emph{Phys. Rev. B}, 72:\penalty0 075443, 2005.
\newblock \doi{10.1103/PhysRevB.72.075443}.
\newblock URL \url{http://doi.org/10.1103/PhysRevB.72.075443}.

\bibitem[Tan et~al.(2007)Tan, Jiang, Huang, Liu, and Hwang]{HTan}
H.~Tan, L.Y. Jiang, Y.~Huang, B.~Liu, and K.C. Hwang.
\newblock The effect of van der waals-based interface cohesive law on carbon
  nanotube-reinforced composite materials.
\newblock \emph{Composites Science and Technology}, 67(14):\penalty0
  2941--2946, 2007.
\newblock \doi{10.1016/j.compscitech.2007.05.016}.
\newblock URL \url{https://doi.org/10.1016/j.compscitech.2007.05.016}.

\bibitem[Mie(1903)]{Mie}
Gustav Mie.
\newblock Zur kinetischen theorie der einatomigen körper.
\newblock \emph{Ann Phys}, 316:\penalty0 657, 1903.
\newblock \doi{10.1002/andp.19033160802}.
\newblock URL \url{http://dx.doi.org/10.1002/andp.19033160802}.

\bibitem[Avendano et~al.(2011)Avendano, Lafitte, Galindo, Adjiman, Jackson, and
  Muller]{Avendano}
C.~Avendano, T.~Lafitte, A.~Galindo, C.~S. Adjiman, G.~Jackson, and E.~Muller.
\newblock Saft-$\gamma$ force field for the simulation of molecular fluids. 1.
  a single-site coarse grained model of carbon dioxide.
\newblock \emph{J Phys Chem B}, 115:\penalty0 11154, 2011.
\newblock \doi{10.1021/jp204908}.
\newblock URL \url{http://dx.doi.org/10.1021/jp204908}.

\bibitem[Kolmogorov and Crespi(2000)]{Kolmogorov}
Aleksey~N. Kolmogorov and Vincent~H. Crespi.
\newblock Smoothest bearings: interlayer sliding in multiwalled carbon
  nanotubes.
\newblock \emph{Phys. Rev. Lett.}, 85(22):\penalty0 4727--4730, 2000.
\newblock \doi{10.1103/PhysRevLett.85.4727}.
\newblock URL \url{http://doi.org/10.1103/PhysRevLett.85.4727}.

\bibitem[Kolmogorov and Crespi(2005)]{Kolmogorov2}
Aleksey~N. Kolmogorov and Vincent~H. Crespi.
\newblock Registry-dependent interlayer potential for graphitic systems.
\newblock \emph{Physical Review B}, 71(23):\penalty0 235415, 2005.
\newblock \doi{10.1103/PhysRevB.71.235415}.
\newblock URL \url{https://doi.org/10.1103/PhysRevB.71.235415}.

\bibitem[Plimpton(1995)]{LAMMPS}
S.~Plimpton.
\newblock Fast parallel algorithms for short-range molecular dynamics.
\newblock \emph{J Comp Phys, 117, 1-19 (1995)}, 117:\penalty0 1--19, 1995.
\newblock URL \url{http://lammps.sandia.gov}.

\bibitem[van Wijk et~al.(2014{\natexlab{b}})van Wijk, Schuring, Katsnelson, and
  Fasolino]{Wijk3}
M.~M. van Wijk, A.~Schuring, M.~I. Katsnelson, and A.~Fasolino.
\newblock Moiré patterns as a probe of interplanar interactions for graphene
  on h-bn.
\newblock \emph{Phys. Rev. Lett.}, 113:\penalty0 135504, 2014{\natexlab{b}}.
\newblock \doi{10.1103/PhysRevLett.113.135504}.
\newblock URL \url{http://dx.doi.org/10.1103/PhysRevLett.113.135504}.

\bibitem[Schuring(2014)]{Annelot}
Annelot Schuring.
\newblock Moiré patterns in graphene in hexagonal substrates.
\newblock \emph{Radbound University Niimegen}, Master Thesis, 2014.

\bibitem[Ng et~al.(2012)Ng, Lau, Bernados-Chamagne, Liu, Sheridan, and Tan]{Ng}
Tuck~Wah Ng, Chun~Yat Lau, Esteban Bernados-Chamagne, Jefferson~Zhe Liu, John
  Sheridan, and Ne~Tan.
\newblock Graphite flake self-retraction response based on potential seeking.
\newblock \emph{Nanoscale Res Lett.}, 7(1):\penalty0 185, 2012.
\newblock \doi{10.1103/PhysRevB.93.235414}.
\newblock URL \url{http://doi.org/10.1103/PhysRevB.93.235414}.

\bibitem[Guerra et~al.(2010)Guerra, Tartaglino, Vanossi, and Tosatti]{Guerra}
Roberto Guerra, Ugo Tartaglino, Andrea Vanossi, and Erio Tosatti.
\newblock Ballistic nanofriction.
\newblock \emph{Nature Materials}, 9:\penalty0 634--637, 2010.
\newblock \doi{10.1038/nmat2798}.
\newblock URL \url{http://doi.org/10.1038/nmat2798}.

\bibitem[Lebedeva et~al.(2012)Lebedeva, Knizhnik, Popov, Lozovik, and
  Potapkin]{Lebedeva}
Irina~V. Lebedeva, Andrey~A. Knizhnik, Andrey~M. Popov, Yurii~E. Lozovik, and
  Boris~V. Potapkin.
\newblock Modeling of graphene-based nems.
\newblock \emph{Physica E}, 44(6):\penalty0 949--954, 2012.
\newblock \doi{10.1016/j.physe.2011.07.018}.
\newblock URL \url{https://doi.org/10.1016/j.physe.2011.07.018}.

\bibitem[Lebedeva et~al.(2011)Lebedeva, Knizhnik, Popov, Lozovik, and
  Potapkin]{Lebedeva2}
Irina~V. Lebedeva, Andrey~A. Knizhnik, Andrey~M. Popov, Yurii~E. Lozovik, and
  Boris~V. Potapkin.
\newblock Molecular dynamics simulation of the self-retracting motion of a
  graphene flake.
\newblock \emph{Phys. Rev. B}, 84:\penalty0 245437, 2011.
\newblock \doi{10.1103/PhysRevB.84.245437}.
\newblock URL \url{https://doi.org/10.1103/PhysRevB.84.245437}.

\bibitem[Jiang(2014)]{JinWuJiang}
Jin-Wu Jiang.
\newblock Registry effect on the thermal conductivity of few-layer graphene.
\newblock \emph{Journal of Applied Physics}, 116:\penalty0 164313, 2014.
\newblock \doi{10.1063/1.4900526}.
\newblock URL \url{http://dx.doi.org/10.1063/1.4900526}.

\bibitem[Baskin and Meyer(1955)]{Baskin}
V.~Baskin and L.~Meyer.
\newblock Lattice constants of graphite at low temperatures.
\newblock \emph{Phys. Rev. B}, 100\penalty0 (2):\penalty0 544, 1955.
\newblock \doi{10.1103/PhysRevB.100.544}.
\newblock URL \url{http://doi.org/10.1103/PhysRevB.100.544}.

\bibitem[Kiang et~al.(1998)Kiang, Endo, Ajayan, Dresselhaus, and
  Dresselhaus]{Kiang}
C.~H. Kiang, M.~Endo, P.~M. Ajayan, G.~Dresselhaus, and M.~S. Dresselhaus.
\newblock Size effects in carbon nanotubes.
\newblock \emph{Phys. Rev. Lett.}, 81(9):\penalty0 1869, 1998.
\newblock \doi{10.1103/PhysRevLett.81.1869}.
\newblock URL \url{http://doi.org/10.1103/PhysRevLett.81.1869}.

\bibitem[Trucano and Chen(1975)]{Trucano}
P.~Trucano and R.~Chen.
\newblock Structure of graphite by neutron diffraction.
\newblock \emph{Nature}, 258:\penalty0 136, 1975.

\bibitem[Gao and Huang(2011)]{Gao}
Wei Gao and Rui Huang.
\newblock Effect of surface roughness on adhesion of graphene membranes.
\newblock \emph{J. Phys. D: Appl. Phys.}, 44:\penalty0 452001, 2011.
\newblock URL \url{http://stacks.iop.org/0022-3727/44/i=45/a=452001}.

\bibitem[Wang et~al.(2012)Wang, Panzik, Kiefer, and Lee]{Wang}
Yuejian Wang, Joseph~E. Panzik, Boris Kiefer, and Kanani K.~M. Lee.
\newblock Crystal structure of graphite under room-temperature compression and
  decompression.
\newblock \emph{Scientific Reports}, 2:\penalty0 520, 2012.
\newblock \doi{10.1038/srep00520}.
\newblock URL \url{http://doi.org/10.1038/srep00520}.

\bibitem[Lynch and Drickamer(1966)]{Lynch}
R.~W. Lynch and H.~G. Drickamer.
\newblock Effect of high pressure on the lattice parameters of diamond,
  graphite, and hexagonal boron nitride.
\newblock \emph{The Journal of Chemical Physics}, 44:\penalty0 181, 1966.
\newblock \doi{http://doi.org/10.1063/1.1726442}.
\newblock URL \url{http://doi.org/10.1063/1.1726442}.

\bibitem[Zhao and Spain(1989)]{ZhaoSpain}
You~Xiang Zhao and Ian~L. Spain.
\newblock X-ray diffraction data for graphite to 20 gpa.
\newblock \emph{Phys. Rev. B}, 40(2):\penalty0 993, 1989.
\newblock \doi{10.1103/PhysRevB.40.993}.
\newblock URL \url{http://doi.org/10.1103/PhysRevB.40.993}.

\bibitem[Clark et~al.(2013)Clark, Jeon, Chen, and Yoo]{Clark}
S.M. Clark, Ki-Joon Jeon, Jing-Yin Chen, and Choong-Shik Yoo.
\newblock Few-layer graphene under high pressure: Raman and x-ray diffraction
  studies.
\newblock \emph{Solid State Communications}, 154:\penalty0 15--18, 2013.
\newblock \doi{10.1016/j.ssc.2012.10.002}.
\newblock URL \url{http://doi.org/10.1016/j.ssc.2012.10.002}.

\bibitem[Hanfland et~al.(1989)Hanfland, Beister, and Syassen]{Hanfland}
M.~Hanfland, H.~Beister, and K.~Syassen.
\newblock Graphite under pressure: Equation of state and first-order raman
  modes.
\newblock \emph{Phys Rev B}, 39:\penalty0 12598, 1989.
\newblock \doi{10.1103/PhysRevB.39.12598}.
\newblock URL \url{http://doi.org/10.1103/PhysRevB.39.12598}.

\bibitem[Lipson and Stokes(1942)]{Lipson}
H.~Lipson and A.R. Stokes.
\newblock The structure of graphite.
\newblock \emph{Proc. Roy. Soc. A}, 181:\penalty0 101--105, 1942.

\bibitem[Lee et~al.(2008)Lee, Lee, Ahn, Kim, Wilson, and John]{LeeLee}
Jae-Kap Lee, Seung-Cheol Lee, Jae-Pyoung Ahn, Soo-Chul Kim, John I.~B. Wilson,
  and Phillip John.
\newblock The growth of aa graphite on (111) diamond.
\newblock \emph{J. Chem. Phys.}, 129:\penalty0 234709, 2008.
\newblock \doi{10.1063/1.2975333}.
\newblock URL \url{http://dx.doi.org/10.1063/1.2975333}.

\bibitem[Dahn et~al.(1990)Dahn, Fong, and Spoon]{Dahn}
J.~R. Dahn, Rosamaria Fong, and M.~J. Spoon.
\newblock Suppression of staging in lithium-intercalated carbon by disorder in
  the host.
\newblock \emph{Phys. Rev. B}, 42:\penalty0 6424, 1990.
\newblock \doi{10.1103/PhysRevB.42.6424}.
\newblock URL \url{https://doi.org/10.1103/PhysRevB.42.6424}.

\bibitem[Norimatsu and Kusunoki(2010)]{Norimatsu}
Wataru Norimatsu and Michiko Kusunoki.
\newblock Selective formation of abc-stacked graphene layers on sic(0001).
\newblock \emph{Phys. Rev. B}, 81:\penalty0 161410, 2010.
\newblock \doi{10.1103/PhysRevB.81.161410}.
\newblock URL \url{http://doi.org/10.1103/PhysRevB.81.161410}.

\bibitem[Charlier et~al.(1991)Charlier, Gonze, and Michenaud]{Charlier3}
J.-C. Charlier, X.~Gonze, and J.-P. Michenaud.
\newblock First-principles study of the electronic properties of graphite.
\newblock \emph{Phys. Rev. B}, 43:\penalty0 4579, 1991.
\newblock \doi{10.1103/PhysRevB.43.4579}.
\newblock URL \url{http://doi.org/10.1103/PhysRevB.43.4579}.

\bibitem[Yoshizawa et~al.(1996)Yoshizawa, Kato, and Yamabe]{Yoshizawa}
Kazunari Yoshizawa, Takashi Kato, and Tokio Yamabe.
\newblock Interlayer interactions in two-dimensional systems: Second-order
  effects causing abab stacking of layers in graphite.
\newblock \emph{J. Chem. Phys.}, 105(5):\penalty0 2099, 1996.
\newblock \doi{http://dx.doi.org/10.1063/1.472076}.
\newblock URL \url{http://dx.doi.org/10.1063/1.472076}.

\bibitem[Charlier et~al.(1994{\natexlab{b}})Charlier, Gonze, and
  Michenaud]{Charlier}
J.~C. Charlier, X.~Gonze, and J.~P. Michenaud.
\newblock First-principles study of the stacking effect on the electronic
  properties of graphite(s).
\newblock \emph{Carbon}, 32 (2):\penalty0 289--299, 1994{\natexlab{b}}.
\newblock \doi{https://doi.org/10.1016/0008-6223(94)90192-9}.
\newblock URL \url{https://doi.org/10.1016/0008-6223(94)90192-9}.

\bibitem[Gao and Tkatchenko(2015)]{WangGao}
W.~Gao and A.~Tkatchenko.
\newblock Sliding mechanisms in multilayered hexagonal boron nitride and
  graphene: The effects of directionality, thickness, and sliding constraints.
\newblock \emph{Phys. Rev. Lett.}, 114:\penalty0 096101, 2015.
\newblock \doi{10.1103/PhysRevLett.114.096101}.
\newblock URL \url{https://doi.org/10.1103/PhysRevLett.114.096101}.

\bibitem[Tao et~al.(2012)Tao, Qing, Yan, and Kuang]{Tao}
Wang Tao, Guo Qing, Liu Yan, and Sheng Kuang.
\newblock A comparative investigation of an ab- and aa-stacked bilayer graphene
  sheet under an applied electric field: A density functional theory study.
\newblock \emph{Chin. Phys. B}, 21:\penalty0 067301, 2012.
\newblock \doi{10.1088/1674-1056/21/6/067301}.
\newblock URL \url{http://doi.org/10.1088/1674-1056/21/6/067301}.

\bibitem[Dolling and Brockhouse(1962)]{Dolling}
G.~Dolling and B.~N. Brockhouse.
\newblock Lattice vibrations in pyrolitic graphite.
\newblock \emph{Phys. Rev.}, 128:\penalty0 1120, 1962.
\newblock \doi{10.1103/PhysRev.128.1120}.
\newblock URL \url{https://doi.org/10.1103/PhysRev.128.1120}.

\bibitem[Lebedev et~al.(2017)Lebedev, Lebedeva, Popov, and Knizhnik]{Lebedeva4}
Alexander~V. Lebedev, Irina~V. Lebedeva, Andrey~M. Popov, and Andrey~A.
  Knizhnik.
\newblock Stacking in incommensurate graphene/hexagonal-boron-nitride
  heterostructures based on ab initio study of interlayer interaction.
\newblock \emph{Phys. Rev. B}, 96:\penalty0 085432, 2017.
\newblock \doi{10.1103/PhysRevB.96.085432}.
\newblock URL \url{https://doi.org/10.1103/PhysRevB.96.085432}.

\bibitem[Shang et~al.(2013)Shang, Cong, Zhang, Xiong, Gurzadyan, and Yu]{Shang}
Jingzhi Shang, Chunxiao Cong, Jun Zhang, Qihua Xiong, Gagik~G. Gurzadyan, and
  Ting Yu.
\newblock Observation of low-wavenumber out-of-plane optical phonon in
  few-layer graphene.
\newblock \emph{J Raman Spec}, 40:\penalty0 70--74, 2013.
\newblock \doi{10.1002/jrs.4141}.
\newblock URL \url{http://doi.org/10.1002/jrs.4141}.

\bibitem[Baranowski et~al.(2013)Baranowski, Mozdzonek, Dabrowski, Grodecki,
  Osewski, Kozlowski, Kopciuszynski, Jalochowski, and Strupinski]{Baranowski}
J.~M. Baranowski, M.~Mozdzonek, P.~Dabrowski, K.~Grodecki, P.~Osewski,
  W.~Kozlowski, M.~Kopciuszynski, M.~Jalochowski, and W.~Strupinski.
\newblock Observation of electron-phonon couplings and fano resonances in
  epitaxial bilayer graphene.
\newblock \emph{Graphene}, 2:\penalty0 115--120, 2013.
\newblock \doi{http://dx.doi.org/10.4236/graphene.2013.24017}.
\newblock URL \url{http://dx.doi.org/10.4236/graphene.2013.24017}.

\bibitem[Lui and Heinz(2013)]{Lui}
Chun~Hung Lui and Tony~F. Heinz.
\newblock Measurement of layer breathing mode vibrations in few-layer graphene.
\newblock \emph{Phys. Rev. B}, 87:\penalty0 121404, 2013.
\newblock \doi{10.1103/PhysRevB.87.121404}.
\newblock URL \url{http://doi.org/10.1103/PhysRevB.87.121404}.

\bibitem[Thornton and Marion(2003)]{Thornton}
Steven~T. Thornton and Jerry~B. Marion.
\newblock Classical dynamics of particles and systems.
\newblock \emph{Fifth Edition}, Brooks Cole, Pacific Grove, CA, 2003.

\bibitem[Tan et~al.(2012)Tan, Han, W.~J.~Zhao, Chang, Wang, Wang, Bonini,
  Marzari, Savini, Lombardo, and Ferrari]{Tan}
P.~H. Tan, W.~P. Han, 1~Z. H.~Wu W.~J.~Zhao, K.~Chang, H.~Wang, Y.~F. Wang,
  N.~Bonini, N.~Marzari, G.~Savini, A.~Lombardo, and A.~C. Ferrari.
\newblock The shear mode of multilayer graphene.
\newblock \emph{Nat. Mater.}, 11:\penalty0 294--300, 2012.
\newblock \doi{http://doi.org/10.1038/nmat3245}.

\bibitem[Cong and Yu(2014)]{Cong}
Chunxiao Cong and Ting Yu.
\newblock Enhanced ultra-low-frequency interlayer shear modes in folded
  graphene layers.
\newblock \emph{Nature Communications}, 5:\penalty0 4709, 2014.
\newblock \doi{10.1038/ncomms5709}.
\newblock URL \url{http://doi.org/10.1038/ncomms5709}.

\end{thebibliography}
